# Using Single Molecule Imaging to Explore Intracellular Heterogeneity

James A. Galbraith* and Catherine G. Galbraith*
*Equal contribution

Oregon Health and Science University
Quantitative and Systems Biology Program in BME and The Knight Cancer Institute
Portland, OR 97239

Address correspondence to galbrcat@ohsu.edu or galbrjam@ohsu.edu

**Abstract**

Despite more than 100 years of study, it is unclear if the movement of proteins inside the cell is best described as a mosh pit or an exquisitely choreographed dance. Recent studies suggest the latter. Local interactions induce molecular condensates such as liquid-liquid phase separations (LLPSs) or non-liquid, functionally significant molecular aggregates, including synaptic densities, nucleoli, and Amyloid fibrils. Molecular condensates trigger intracellular signaling and drive processes ranging from gene expression to cell division. However, the descriptions of condensates tend to be qualitative and correlative. Here, we indicate how single-molecule imaging and analyses can be applied to quantify condensates. We discuss the pros and cons of different techniques for measuring differences between transient molecular behaviors inside and outside condensates. Finally, we offer suggestions for how imaging and analyses from different time and space regimes can be combined to identify molecular behaviors indicative of condensates within the dynamic high-density intracellular environment.

1. Introduction

The idea that interactions between individual members drive a collective output is fundamental to biology. It is how fish, birds, and migrating cells only interact with their proximal neighbors yet create an emergent behavior for the school, flock, or sheet of cells (Friederici, 2009; Friedl and Gilmour, 2009). There is a tremendous amount of excitement that similar principles may apply to molecules within the cytoplasm – groups of molecules work together within the cytoplasm to form functionally significant molecular aggregates or condensates. However, how condensates give rise to cellular behaviors remains to be determined. It is generally acknowledged that the field "desperately needs" new tools (Leslie, 2021). Proposals have been made suggesting using single-molecule (SM) super-resolution microscopy and single-particle tracking, but it is unclear how they should be applied (Lyon et al., 2021). Here, we explore how these tools can be used to identify and quantify the heterogeneous molecular behaviors indicative of condensate formation.

Condensates frequently arise from liquid-liquid phase separations (LLPS), transforming a single, uniform liquid phase into two compositionally distinct liquid phases (Lafontaine et al., 2021). The concept of LLPSs rose to prominence with the discovery of the liquid-like behavior of *Caenorhabditis elegans* P granules, a class of perinuclear RNA granules specific to the germline (Brangwynne et al., 2009). P granules are separated from the rest of the cytoplasm, but they are not confined by a membrane, giving rise to the term "membrane-less" organelle. Since an LLPS can merge with another LLPS and coalesce, similar to the everyday example of oil droplets separating from water when a bottle of salad dressing is shaken, this visual imagery has led to the observational properties of droplet roundness and merging becoming de facto standard metrics for describing LLPSs (McSwiggen et al., 2019b). Using descriptors such as fusion and dispersion as indicators of LLPSs has resulted in the current qualitative state of the field, which has been documented by an in-depth review of 33 papers reporting on the properties of LLPSs (McSwiggen et al., 2019). Adding to the challenges of quantifying condensates by describing them simply as droplets is the production of unwanted puncta when tools like 1,6-hexanediol are used to disrupt the hydrophobic interactions within LLPSs (Leslie, 2021).

The term condensate has recently been extended to include functional molecular aggregates or heterogeneities, such as synaptic densities, nucleoli, and membrane clusters, which are not LLPSs (Banani et al., 2017). As the number of identified condensates expands, so does the recognition of their importance in normal physiology and pathologies ranging from cancer to Alzheimer's (Lyon et al., 2021). Condensates now appear almost everywhere and are involved in nearly everything. However, there is still no consensus on how they should be quantified. The discord arises from the experimental difficulties in measuring the behaviors of molecules that move with instantaneous velocities of ~10m/s in the cytoplasm (Bray, 2000). Molecules rotate around and change axes so fast that a single molecule will meet with every other molecule inside a bacterium cell every second (Milo and Phillips, 2015). While this environment facilitates the formation of the transient molecular aggregations that underlie many condensates, it also makes it more challenging to quantify them.

2. **Quantifying the molecular population**

Many different microscopy techniques have been used to identify condensates (Zhang et al., 2023). While techniques such as widefield transmitted light and fluorescent confocal provide shape descriptors about droplets, the "gold standard" for measuring what happens inside a condensate has been fluorescence recovery after photobleaching (FRAP) (McSwiggen et al., 2019b). During FRAP, a region in a condensate expressing a fluorescent protein is photobleached. The fluorescence recovery into the bleached region of interest (ROI) is recorded and then fit to a model that infers the underlying molecular dynamics (Fig 1a). Practically, the bleach ROI must be easily distinguished from the surrounding environment and be significantly less than the size of the condensate to avoid "edge" effects. When a conventional microscope such as confocal is used, this sets a lower limit on the ROI size as the diffraction limit of 250 µm. It is important that the condensate does not rapidly change size or position during the bleach or recovery time to avoid motion artifacts or ambiguous localization. It is also important to remember that typical analyses assume that the measurement can represent a combination of fast and slow-moving molecules but the mixture is spatially homogeneous. While FRAP measures dynamics only in the bleached region, it does not represent the exchange between the dilute and dense phases of an LLPS. Thus, FRAP cannot be used to prove that a liquid-like state has arisen through an LLPS process (Alberti et al., 2019). Ignoring these underlying FRAP principles has yielded literature values for the mobility of the same molecular species varying by up to three orders of magnitude (Taylor et al., 2019).

Another standard tool used to quantify condensates is fluorescence correlation spectroscopy (FCS), which records the fluctuations in fluorescent signal as molecules pass through an illuminated spot using specialized photon counting hardware (Fig 1b). FCS is also an ROI technique that provides a measure of population mobility. However, FCS measures the mobility of individual molecules so that it can record faster dynamics than FRAP. In addition, the FCS ROI can be made smaller than the resolution of conventional microscopes by using stimulated emission depletion (STED) optics (Honigmann et al., 2013). While FCS data is frequently modeled to account for multiple molecular species and interactions, FCS, like FRAP, reports the average behavior of the individual molecules in the ROI (Schwille, 2001). However, FCS requires lower-density sample labeling to measure individual molecules' high-speed behaviors than FRAP. The differences between these techniques illustrate that higher molecular density information comes at the price of slower speed, and neither technique can report the behavior of the individual molecules in the condensate (Fig 1c).

### 3. Quantifying single-molecules behavior

The ability to control the ON/OFF state of the fluorescent label and the development of SM super-resolution microscopy dramatically increases the opportunities to study the behaviors of individual molecules in crowded environments. By making only a subset of molecules visible within a given camera exposure, the fluorescence emission of the ON molecules can be fit to a Gaussian profile, which approximates the shape of the point spread function from the emission of a single point of light (Betzig et al., 2006; Hess et al., 2006; Rust et al., 2006; van de Linde et al., 2011) (Fig 2a). The fit's accuracy is called the localization precision and depends upon the fluorophore's brightness and background noise (Thompson et al., 2002). Experimentally, fits are often within 10-20 nanometers, which provides near molecular precision.

How many molecules can be imaged with SM resolution at any given time within a condensate? Theoretical calculations suggest condensates can be as small as 10 nm (Forman-Kay et al., 2022). So, an estimate of the number of molecules that can be identified inside a condensate during a single snapshot is the number of non-overlapping circles of diameter equal to the molecular localization precision (Specht, 2021). Obviously, the higher the localization precision, the more molecules will fit inside a given size condensate. As shown in Fig. 2b, this relationship follows a nominal second-order power law. While the calculation assumes that all molecules are immobile, fully labeled, and activated during a single acquisition frame, an obvious oversimplification, it is an informative guide to consider when developing parameters for imaging and quantitative analyses.

**3.1 Fluorescent labels and the photon budget**

Imaging begins with labeling the molecules, and the label chosen will significantly impact the dynamics that can be quantified. Proteins in live cells are typically labeled with genetically expressed photoactivatable fluorophores (e.g., Dendra2, mEos3.2) (Pakhomov et al., 2017; Zhang et al., 2012) or brighter organic dyes such as the JFdyes (Grimm et al., 2020), their variants, photoactivatable PA-JF (Grimm et al., 2016) and more photostable JFX dyes. Organic dyes can label proteins expressing peptide tags such as SnapTag (Keppler et al., 2004a; Keppler et al., 2004b) or HaloTag (Los et al., 2008), but care must be taken to wash any free unlabeled dye out of the cell to avoid untagged fluorophores being analyzed and contributing to fluorescent background, which will reduce localization precision. It is also essential to consider that what seems like minor changes to the labeling strategy, such as labeling a different heterodimer subunit, can subtly change the protein conformation, causing a protein that localizes properly and interacts with known binding partners to have a slightly different affinity for ligand (Galbraith et al., 2018). Therefore, maintaining endogenous expression levels by CRISPR-Cas9 genome editing and demonstrating that the label tag does not induce functional artifacts when measuring molecular behaviors involved in the clustering and phase separations are prime considerations at the start of the experimental design.

In addition to brightness, it is important to consider a fluorophore's "photon budget." Each fluorophore has only a finite number of photons to release. How the fluorophore releases those photons depends on its quantum yield and the photobleaching rate. Quantum yield is the number of photons emitted per photon absorbed, and the photobleaching rate is how many exposures one can acquire before the signal-to-noise ratio is unacceptable for quantification. The two parameters are not independent. Increasing laser power will increase the number of photons emitted during a single camera exposure and increase localization precision. However, it also depletes the pool of available photons more quickly and decreases the number of frames that a molecule can be imaged and, hence, the total length of the molecular trajectories. Thus, to answer the biological question of interest, the labeling choice must consider the molecular copy number, the labeling density required for the imaging technique, the relative ease of ON/OFF control among multiple colors, and the amount of time a given molecule must be followed.

Since imaging parameters can affect fluorophore output, they should be judiciously adjusted to match molecular speeds. For example, lower laser powers will slow the release of photons, prolonging the time until a fluorophore is completely bleached. However, this will require longer exposure times to collect the same strength signal and achieve the same localization precision. If the camera exposure is too long, the moving molecule will appear blurred due to the molecule moving a distance greater than the number of camera pixels that define the localization precision. A useful experimental protocol is to add a "rest" period between rapid camera exposures to increase fluorophore lifetime, but if the molecules move too quickly, then it may be difficult to "connect the dots" between consecutive frames during SMT. Despite needing to consider multiple parameters, carefully balancing laser pulse width and shorter exposure times with dark intervals can extend the fluorophore lifetime and stretch the photon budget to quantify dynamics accurately (Elf et al., 2007).

### 3.2 Testing that the budget was spent wisely

Since these labeling, laser power, and camera acquisition rate choices are based on deductive reasoning; it is prudent to use preliminary experimental data to determine whether the localization precision, the molecular density, and the length of the molecule trajectories are sufficient or if further imaging optimization is needed. For this analysis, there are many localization and tracking software packages available. We refer the reader to a recent comparison challenge (Sage et al., 2019) and a newer combined tracking and analyses package (Kuhn et al., 2021). However, there are some underlying principles that should be considered when choosing a package. Since single molecule tracking (SMT) was initially developed for sparsely labeled samples (Geerts et al., 1987; Qian et al., 1991), conventional nearest neighbor tracking works most reliably if the density is low. Higher-density tracking requires the algorithms to incorporate a predictive capability (Jaqaman et al., 2008). But, at too high a density, the paths of too many ON molecules are likely to cross between camera exposures, making it difficult for even predictive SMT algorithms to untangle the tracks accurately (Jaqaman et al., 2016). While parameters relating to the fit of the molecule to the theoretical Gaussian can be used to aid in identification, entanglement can still lead to ambiguous tracking. Thus, the density and temporal resolution of the population need to be balanced between the number and speed of movement of individual molecules.

### 3.3 Quantifying the trajectories

Additional analysis is needed to turn tracks into quantitative metrics (Fig 2c-d). One of the most common is calculating the diffusion coefficient using the mean square displacement (MSD) (Pandey et al., 2023). When plotted on a log-log scale, MSD vs. time lag ($\Delta t$) is linear for pure diffusion, with the slope yielding the diffusion coefficient (Fig 2c). However, the MSD is very sensitive to experimentally induced inaccuracies that arise with rapidly diffusing molecules, including short trajectories and complex multi-state behaviors, making it more error-prone for quantifying condensates. Since motion in a crowded environment is often sub-diffusive (power law with a negative exponent) or super-diffusive (power law with a positive exponent), momentum scaling spectrum analysis (MSS), which extends the analysis to other moments of displacement beyond the second-order moment (the MSD), can be more accurate (Ewers et al., 2005).  However, it is important to determine if the biological question that needs to be answered requires calculating complex high-speed individual stochastic behaviors or if the need is to separate functionally significant molecular heterogeneity from stochastic "noise."  If the dynamics of the individual molecules are not needed, an alternative analysis approach is to use the individual trajectories to fit time-dependent displacement histograms with Gaussian probability density mixture models and obtain the diffusion coefficient and fraction of particles in each state (Fig 2d).  Even if all the particles are not purely diffusing, unlike FRAP and FCS, these models can still infer transition states, which can be important for deciding if the local molecular behaviors are indicative of a condensate (Boka et al., 2021).

Another analytical approach is to use the relative angles of trajectory motion.  Purely diffusing molecules explore in all directions, exhibiting an isotropic distribution of angles. In contrast, confined molecules, such as those in a condensate, tend to revisit the same space, with a 180° bias, bouncing back and forth between boundaries.   Trajectories can be sorted into populations by either average displacement or another classifier, and the angles are plotted for each population (Fig 2d).  A useful metric to include for transiently interacting molecules is the mean first-passage time (MFPT) for a particle to reach a target.  It is calculated as the negative time derivative of the survival probability of the particle in the searching state (Condamin et al., 2008; Polizzi et al., 2016).

Molecular interactions or individual molecule dwell times can also be calculated.  They are based on trajectory lengths and combined into a survival probability distribution, which can be fitted with a two-component exponential decay model for each component (Spot-On) (Hansen et al., 2018), variational Bayesian statistics (vbSPT) (Persson et al., 2013), or non-parametric Bayesian statistics (SMAUG) (Karslake et al., 2021). Alternatively, SMT can also be used to calculate the off rates or residence times of two molecules interacting. Significantly, the inherent space-time imaging tradeoff between capturing single molecule and population behaviors can be used to create a bias for seeing these residence times -- long exposures will blur the fast-moving molecules and allow lower laser powers to yield longer trajectories on the slower-moving molecules (Boka et al., 2021).

When SM imaging is used only to localize the static position of molecules, it has revealed that molecular condensates can be heterogeneous, composed of sub-compartments and structured elements with heterogenous molecular behaviors (Niewidok et al., 2018). Multicolor SM super-resolution has resolved the nanoarchitecture of interacting proteins within a condensate in fixed cells (Kanchanawong et al., 2010). SM super-resolution has also identified patterns of molecular entry and exit to condensates in live cells (Shroff et al., 2008). Analysis of the distribution of localized molecules typically involves grouping molecules according to their centroid, distribution, density, or hierarchy (Fig 2e). Popular approaches include density-based spatial clustering with noise (DBSCAN) (Ester et al., 1996) or Voronoi tessellation (Levet et al., 2019), and extensive lists of clustering algorithms (and their applicability to a specific condition) can be found elsewhere (Xu and Tian, 2015). It is important to note that obtaining high spatial density requires stochastically turning on a different subset of non-overlapping molecules to build up high molecular density super-resolution images over much longer acquisition times than the spatially sparse high-speed data obtained from SMT (Fig 2f).

## 4. Putting it all together

Visualizing and quantifying heterogeneous molecular behaviors within dense molecular fields is always a space-time trade-off. Capturing the high-speed movement of individual molecules requires a sparse enough field so that the molecular trajectories are not entangled between successive camera frames. Information about the dynamics of the individual comes at the price of the ensemble's dynamics; capturing the dense ensemble's behavior often means sacrificing the behavior of the individual molecules. However, SM approaches offer the advantage of an imaging field that is typically not restricted to a small ROI and thus includes both condensates and their surrounding context. Analyses can be readily devised to compare molecular dynamics inside and outside an aggregate to identify the existence of a condensate. Additionally, the differences between techniques make a strong case for combining multiple approaches to characterize the molecular organization of condensates (Table 1).

An excellent example of using multiple space and time imaging and analysis techniques to determine the characteristics of a biomolecular condensate and how it functions can be found in the examination of Herpes Simplex Virus I viral replication compartments (RCs) (McSwiggen et al., 2019a). RCs grow over the course of infection and exhibit characteristics of liquid drops, such as fusion and a spherical shape. They are also enriched in RNA Polymerase II (Pol II) proteins with intrinsically disordered domains (IDRs). FRAP experiments suggested that Pol II was incorporated and sequestered with the RC. Thus, by "standard metrics," RCs seem to be LLPSs. However, 1,6-hexanediol, which disrupts interactions between the IDRs, did not inhibit Pol II enrichment in RCs. Labeling HALO-Pol II with equal amounts of JF549 and PA-JF646 and using the JF549 signal as a diffraction-limited mask for RCs and the PA-JF646 for single molecule tracks enabled SMT inside vs. outside the RCs and yielded the unexpected result that the diffusion coefficient was the same inside and outside the RCs. If RCs were a separate phase, one would expect a different diffusion coefficient. In addition, Pol II molecules freely diffused in and out of the RC; there was no physical or energetic barrier hindering motion despite nearly 70% of the molecules being in a bound state inside the RC. However, when analyzed at the single molecule level, it was revealed that the majority of Pol II was bound even when transcription was inhibited and formed hubs because of the virus' genome accessibility within RCs at multiple length scales. The data suggested a conclusion that contradicts the FRAP results -- in the case of TCs, LLPS is not necessary to drive local hub formation and functional compartments (McSwiggen et al., 2019a).

In summary, understanding the exciting biology of molecular condensates is complicated by the technical challenges in seeing transient molecular inhomogeneities that move around within the cell and are too small to detect with conventional microscopy (Villegas et al., 2022). As we learn more about condensates, it becomes clear that even within condensates, there are additional functionally significant heterogeneities in the behaviors of the molecules. Thus, in addition to applying biological perturbations, in-vitro studies (Case et al., 2022), and modeling (Laghmach and Potoyan, 2021; Shimobayashi et al., 2021), we need to consider applying multiple techniques to quantify single-molecules as well as populations of molecules (Munoz-Gil et al., 2022; Wu et al., 2019). No one technique will provide a complete picture (Table 1). So, just as we use multiple biological techniques to answer a question, we must also use multiple imaging techniques targeting different lengths and time scales to map the molecular choreography of the dynamic cytoplasm.

**ACKNOWLEDGEMENTS**

Financial support was provided by a grant from the W. M. Keck Foundation and NIH grant 1R01GM117188.

**FIGURES**

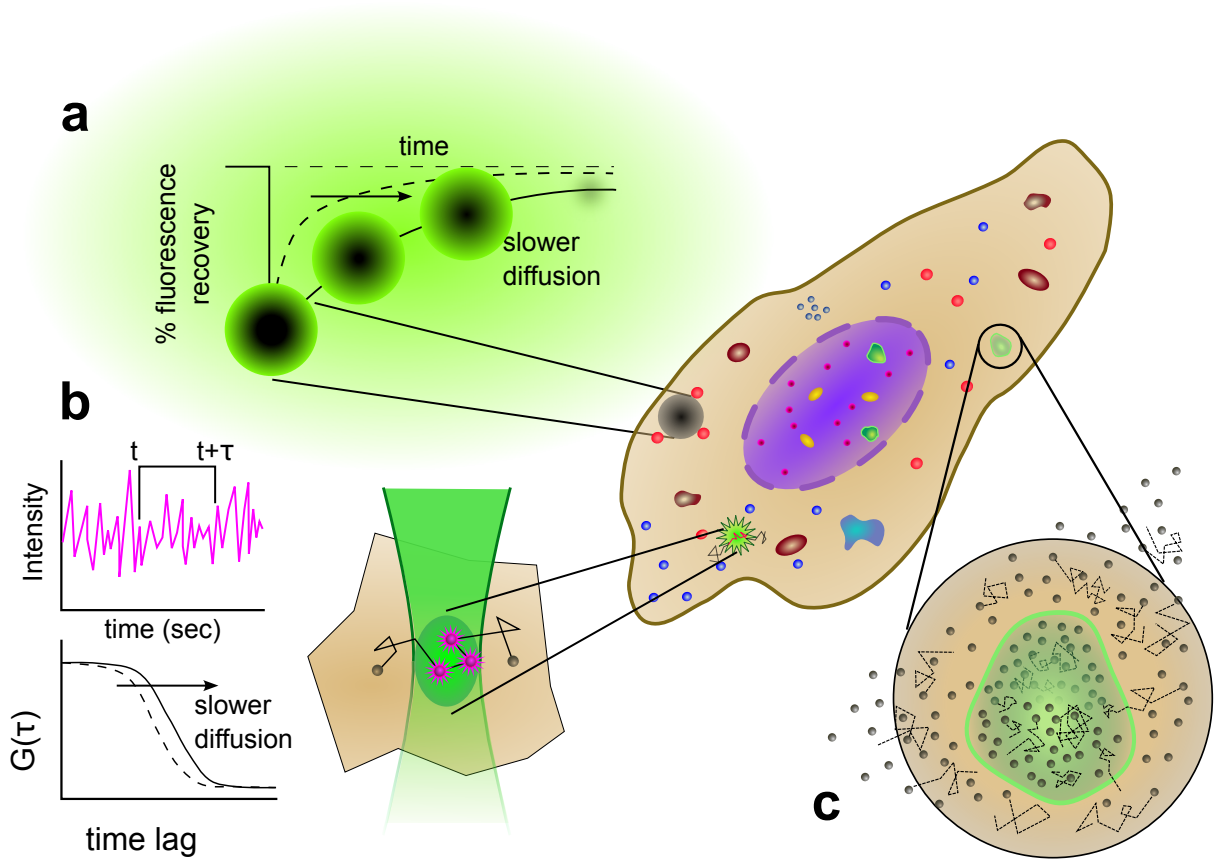

**Figure 1. Ensemble techniques for quantifying the behaviors of fluorescent molecules**. a) Fluorescence Recovery After photobleaching (FRAP) bleaches a region of interest (ROI) and fits the recovery of fluorescence in the bleached region to describe the molecular behavior of the population. b) Fluorescence Correlation Spectroscopy (FCS) monitors the fluctuations in the fluorescence of individual molecules within an excitation volume and uses temporal autocorrelation to quantify average molecular behaviors. c) However, condensates are often crowded molecular environments with heterogeneous, not homogeneous, molecular behaviors that can change over time.

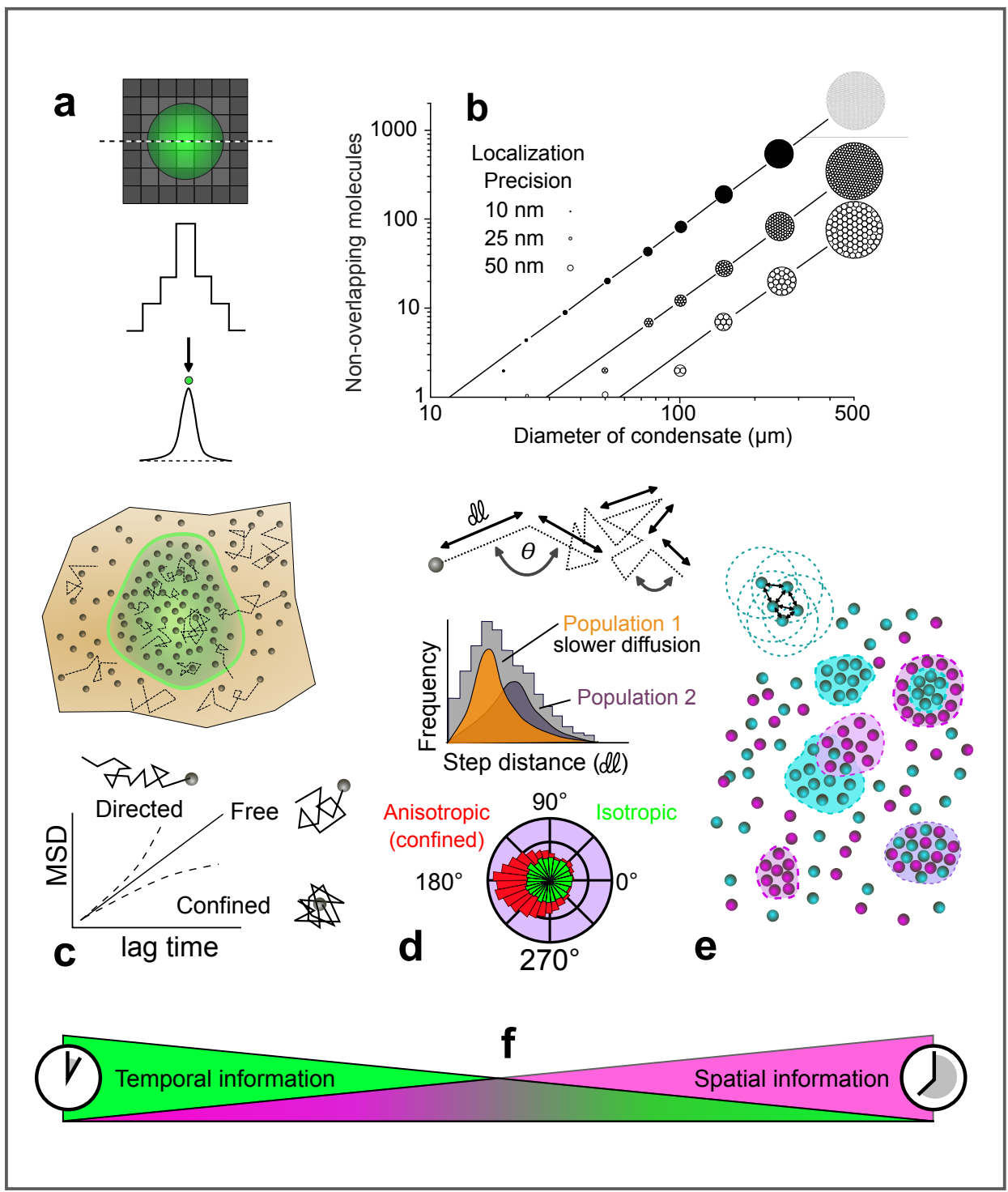

**Figure 2. Single-molecule techniques for quantifying molecular behaviors.** a) By controlling the fluorescent ON/OFF state so that only a small subset of molecules is visible at any given time, the position of the ON molecules can be mathematically determined (localized)

with high precision. b) When the distance separating each visualized molecule equals the localization precision, the maximum number of non-overlapping "molecules" that will fit inside a circular region follows a second-order power law, giving a rough upper bound on molecular density.  c) The trajectories or positions of individual molecules can be spatially separated according to whether they are inside or outside a condensate. Trajectories obtained by single-molecule tracking (SMT) can be analyzed by techniques such as mean squared displacement (MSD) to biophysically quantify their diffusivity and type of motility (directed, free, or confined).  d) The distribution of single molecule trajectory length or orientation angle can be used to identify different diffusive behaviors in populations of molecules or whether molecule movements are anisotropic and confined.  e) Localizing many stationary or slowly moving molecules can define molecular nanoarchitecture and quantify distributions of molecular clusters over a larger cell region but requires a longer collection window.  f) The density of spatial information is inversely proportional to collection time, resulting in a trade-off between spatial and temporal resolution of the final dataset.  Fast events are blurred at long times, while at short times requiring higher excitation power, molecules are bleached faster, yielding fewer trajectories.

**Table 1. Comparison of Ensemble and Single Molecule Analyses.** The color indicates how effective the technique is for providing the indicated measurements, with green good, red poor, and yellow fair.

| | | Method | Spatial Inhomogeneities | Individual molecule behavior | Population dynamics | Quantifies | Comments |
|---|---|---|---|---|---|---|---|
| ENSEMBLE | Measurements only report behavior in ROI. | FRAP | 🔴 | 🔴 | 🟢 | Bulk diffusion | Assumes homogeneous mix of behaviors |
| | | FCS | 🔴 | 🔴 | 🟢 | Population kinetics and binding based on defined model | Measures molecules but does not report data for individuals |
| SINGLE MOLECULE | Whole field measurements, which can be analyzed within local subregions. Needs to be repeated many times to increase spatial density. | MSD | 🟢 | 🟢 | 🟡 | Individual molecular biophysical behaviors | Can be noisy and tracking may be ambiguous at high molecular density |
| | | Distribution of pooled track lengths | 🟢 | 🔴 | 🟡 | Subpopulation kinetics | Better at uncovering multiple states |
| | | Distribution of pooled track angles | 🟢 | 🔴 | 🔴 | Anisotropy: Confinement and subdiffusive behavior | Lose identity of individual molecules |
| | | Cluster | 🟢 | 🔴 | 🔴 | Structural distribution of molecules | Many acquisition cycles for high-structural resolution |